\documentclass[10pt,aps,pre,twocolumn,superscriptaddress,floatfix,longbibliography]{revtex4-1}
\usepackage{graphicx}
\usepackage{amsmath}
\usepackage{amssymb}
\usepackage{hyperref}
\graphicspath{{figs/}}

\DeclareMathOperator{\Prob}{Prob}
\DeclareMathOperator{\Var}{Var}
\begin{document}
\title[Anisotropy in electrical conductivity]{Anisotropy in electrical conductivity and transparency of films of aligned conducting rods}

\author{Nikolai~I.~Lebovka}
\email[Corresponding author: ]{lebovka@gmail.com}
\affiliation{Department of Physical Chemistry of Disperse Minerals, F.\,D.\,Ovcharenko Institute of Biocolloidal Chemistry, NAS of Ukraine, Kiev, Ukraine, 03142}
\affiliation{Department of Physics, Taras Shevchenko Kiev National University, Kiev, Ukraine, 01033}

\author{Yuri~Yu.~Tarasevich}
\email[Corresponding author: ]{tarasevich@asu.edu.ru}
\affiliation{Laboratory of Mathematical Modeling, Astrakhan State University, Astrakhan, Russia, 414056}

\author{Nikolai~V.~Vygornitskii}
\email{vygornv@gmail.com}
\affiliation{Department of Physical Chemistry of Disperse Minerals, F.\,D.\,Ovcharenko Institute of Biocolloidal Chemistry, NAS of Ukraine, Kiev, Ukraine, 03142}

\author{Andrei~V.~Eserkepov}
\email{dantealigjery49@gmail.com}

\author{Renat~K.~Akhunzhanov}
\email{akhunzha@mail.ru}
\affiliation{Laboratory of Mathematical Modeling, Astrakhan State University, Astrakhan, Russia, 414056}

\date{\today}

\begin{abstract}
Numerical simulations by means of the Monte Carlo method  have been performed to study the electrical properties of a two-dimensional composite filled with rodlike particles. The main goal was to study the effect of the alignment of such rods on the anisotropy of its electrical conductivity. A continuous model was used. In this model, the rods have zero-width (i.e. infinite aspect ratio) and they may intersect each other. To involve both the low conductive host matrix, and highly conductive fillers (rods) in the consideration, a discretization algorithm based on the use of a supporting mesh was applied. The discretization is equivalent to the substitution of rods with the polyominoes. Once discretized, the Frank--Lobb algorithm was applied to evaluate the electrical conductivity. Our main findings are (i) the alignment of the rods essentially affects the electrical conductivity and its anisotropy, (ii)the  discrete nature of computer simulations is crucial. For slightly disordered system, high electrical anisotropy was observed at small filler content, suggesting a method to enable the production of optically transparent and highly anisotropic conducting films.
\end{abstract}

\maketitle

\section{Introduction}\label{sec:intro}

Thin films filled with highly conductive rodlike  particles  (such as carbon nanotubes, metal nanowires, etc.) are of increasing interest, particularly, for the production of flexible transparent conductors (for review, see, e.g.,~\cite{Hecht2011} and the references therein). Transparent conducting electrode devices find diverse applications in solar cells, touch-screens and transparent heaters~\cite{De2010,Mutiso2013, Ackermann2016,Kumar2016,Kumar2017}. These promising applications are inspiring both theoretical and experimental studies in this field~\cite{Taherian2016}.

Geometrical percolation is defined as the formation of a connected cluster of particles that spans the whole system~\cite{Stauffer1992}. In the different models, it may be assumed that the particles can either penetrate each other (soft overlapping particles), or may contain both internal ``hard cores'' (an impenetrable part of the particle) and external ``soft shells'' responsible for charge transfer~\cite{Balberg1987}. It is useful to introduce a number density term, $n$, defined as the number of particles per unit volume for three-dimensional (3D) systems or per unit area for two-dimensional (2D) systems. For randomly distributed, overlapping particles, the total fraction of the space covered by the particles (the filling fraction) can be evaluated as in~\cite{Mertens2012}
\begin{equation}\label{eq:phi}
p=1-\exp(-nV_p),
\end{equation}
where $V_p$ is the volume or area of each individual particle for the 3D or 2D problems, respectively.

Percolation problems for elongated particles have been extensively studied both in 3D and 2D geometries. Investigated 3D geometries include randomly oriented overlapping ellipsoids~\cite{Garboczi1995,Yi2004}, overlapping capped cylinders~\cite{Balberg1984,Balberg1984a,Balberg1987,DeBondt1992},  soft-core capped cylinders~\cite{Berhan2007} and soft-core prisms~\cite{Saar2002}. For 2D percolation, randomly oriented overlapping ellipses~\cite{Xia1988,Yi2002}, and rectangles~\cite{Li2013} have also been considered. Analytical approximations and Monte Carlo (MC) simulations have been applied to derive the percolation thresholds for overlapping ellipsoids~\cite{Yi2004} and
overlapping ellipses~\cite{Yi2002}.

The most popular models for describing percolation problems have been the excluded volume (3D) or excluded area (2D) approaches. It has been conjectured that the number density at the geometrical percolation threshold, $n_c$, is inversely proportional to the excluded volume (3D) or excluded area (2D) of the particles, $V_{ex}$,~\cite{Balberg1984a}
\begin{equation}\label{eq:Inverse}
%n_c \propto 1/V_{ex}.
n_c \propto V^{-1}_{ex}.
\end{equation}

The value of $V_{ex}$ depends upon the system dimensionality, the geometry of the particles and their relative orientations.

In 3D systems for oriented cylinders of length $l$ and diameter $d$, the excluded volume $V_{ex}$ is given by
\begin{equation}\label{eq:Vex3D}
V_{ex}=2d l^2\left[\langle \sin\theta \rangle+\pi/a+(2\pi/3)/a^2\right],
\end{equation}
where $\theta$ is the angle between two particles, $a=l/d$ is the aspect ratio, and $\langle \cdot\rangle$ corresponds to the number-averaged value. For random orientations of cylinders,  $\langle \sin\theta \rangle=\pi/4$, and, in the limit of $a\rightarrow \infty$ (slender-rod limit), we have $V_{ex}=(\pi/2)d l^2$.

In the slender-rod limit, the excluded volume rule for capped cylinders was confirmed by cluster expansion approach~\cite{Bug1985} and MC simulations~\cite{Neda1999}. However, for a finite aspect ratio, important deviations from this rule were observed~\cite{Neda1999,Florian2001}.
From the numerical data of simulations~\cite{Berhan2007} the following general expression for $n_c$ was derived
\begin{equation}\label{eq:3dnc}
n_c = \left(1+ c_1 a^{-c_2}\right)V^{-1}_{ex},
\end{equation}
where $c_1=3.526$ and $c_2=0.569$.

The effects of partial penetration on percolation behavior  have also been analyzed~\cite{Balberg1987,DeBondt1992,Berhan2007}. The degree of penetration was defined by the ratio of the diameter of the core to the outer diameter of the soft shell, $\gamma$~\cite{Berhan2007}. For the soft-core limit $\gamma=0$ and for the hard-core $\gamma=1$. An inverse proportionality rule (Eq.~\eqref{eq:Inverse}) was also observed for this hard core/soft shell model. For the slender-rod limit ($a\rightarrow \infty$), the constant of proportionality was shown to be dominated by the aspect ratio $a$, with there being little dependence  on the $\gamma$ value. The effects of size dispersity on electrical percolation in rod networks have also been analyzed~\cite{Otten2009,Otten2011,Mutiso2012}.
The impact of the orientation of the cylinders on percolation behavior was discussed in~\cite{Munson1991}. A detailed review of the effects of aspect ratio, dispersity in filler size and electrical properties, orientation, flexibility and waviness on the percolation behavior of systems of rodlike particles has recently been presented by~\cite{Mutiso2015}.

In 2D systems for oriented capped rectangles of length $l$ and width $d$, the value of  $V_{ex}$ can be derived from the formulae presented in~\cite{Balberg1984a}
\begin{equation}\label{eq:Vex2D}
V_{ex}=l^2\left[\langle \sin\theta \rangle+4/a+\pi/a^2\right].
\end{equation}
For random orientations of rectangles,  $\langle \sin\theta \rangle=2/\pi$, and in the limit of $a\rightarrow \infty$ we have $V_{ex}=(2/\pi)l^2$.

The following interpolation formulas for the total critical filling fraction of particles at the percolation points have been obtained for randomly oriented ellipses~\cite{Xia1988}
\begin{equation}\label{eq:ellipses}
p_c= 1- (1 + 4y)/(19 + 4y),
\end{equation}
and rectangles~\cite{Li2013}
\begin{equation}\label{eq:rectangles}
p_c= 1-(1.28 + y)/(6.73 + y).
\end{equation}
Here, $y = a +1/a$; where $a$ is the aspect ratio.

Considerable attention has been paid to the 2D percolation problem of  zero-width rods ($a=\infty$)~\cite{Pike1974,Balberg1983,Robinson1983,Robinson1984,Li2009,Dalafi2010,Mertens2012,Mietta2014}. An inverse proportionality rule $n_c\propto V^{-1}_{ex}\propto l^{-2}$~\eqref{eq:Inverse} has also been confirmed for this system. For randomly oriented rods, Pike and Seager using a MC simulation obtained $n_cl^2\simeq 5.71$~\cite{Pike1974}. Subsequently, for $n_cl^2$, the following values have been obtained   $5.615$~\cite{Balberg1983}, $5.63$~\cite{Dalafi2010}, $ 5.6372858(6)$~\cite{Mertens2012}, $5.754$~\cite{Mietta2014} and $5.63726\pm 0.00002$~\cite{Li2009}. The total excluded area for this problem can be estimated as $V_{ex}^t=n_cV_{ex}=(2/\pi)n_c l^2\approx 0.64n_c l^2$.

The more general case of an anisotropic system of rods has also been analyzed~\cite{Balberg1983,Balberg1983a,Balberg1984}. The rods were aligned with respect to a selected direction, $x$. Their axes were randomly distributed within some  interval such that, $-\theta_m \leq \theta\leq \theta_m$, where $\theta_m\leq \pi/2$. The isotropic case is given by $\theta_m=\pi/2$, and the smaller the $\theta_m$ the higher the degree of orientation. The macroscopic anisotropy of the system was characterized by the value
\begin{equation}\label{eq:Anisotropy}
A = \langle |\cos\theta| \rangle/\langle |\sin\theta| \rangle = \sin \theta_m/(1-\cos \theta_m).
\end{equation}
Here and below $A=P_{\parallel}/P_{\perp}$ to simplify notation.
For the isotropic case, $A=1$, whereas for the highly anisotropic case ($\theta_m\rightarrow 0$),  $A\rightarrow \infty$~\cite{Pike1974,Balberg1982a}.

The percolation threshold for this system increases with the macroscopic anisotropy $A$~\cite{Balberg1983a,Balberg1984a}. The dependence of the percolation number density on the degree of macroscopic anisotropy was approximated as
\begin{equation}\label{eq:NumberDensityAnisotropy}
n_c^*=n_c/n_c^i=0.5(A + A^{-1}),
\end{equation}
where $n_c^i$ is the critical density of rods for the isotropic case.

Transport properties and in particular the electrical conductivity of systems filled with anisotropic conducting particles are closely related to geometric percolation behavior. The electrical percolation can reflect not only the geometric connectivity of the particles but also aspects of the charge transfer mechanism between individual particles~\cite{Berhan2007}. The conductivity exponent $t$ for the 2D percolation of rods has been discussed in several works~\cite{Balberg1982,Balberg1983,Hecht2006,Li2010,Zezelj2012}. Experimental studies on the electrical conductivity in a composite of conducting rods gave the following interval for the conductivity exponent, $1.5<t<2.1$~\cite{Balberg1982}. Balberg et al. obtained a value of $t=1.24\pm0.03$ by analyzing the conductivity of a percolation cluster of rods~\cite{Balberg1983}. On the contrary, for the situation with differences between the resistances inside the rod, $R_s$, and at the rod-rod junctions, $R_j$, it was proposed that the conductivity exponent $t$ can vary from $0$ when $R_j\ll R_s$ to the universal value of lattice percolation, $t\simeq 1.3$ when $R_j\gg R_s$~\cite{Hecht2006}. Subsequently, MC simulations indicated that the $t$ value extracted from the size-dependent conductivities of systems exactly at the percolation threshold was independent of $R_j/R_s$ ($t=1.280\pm 0.014$)~\cite{Li2010}. However, the $t$ value extracted from the density-dependent conductivities of systems well above the percolation threshold were significantly dependent on $R_j/R_s$. A numerical MC study of the conductivity of random rod networks for a wide range of densities and junction-to-rod conductance ratios has been performed~\cite{Zezelj2012}. Three limiting cases with different conductivity exponents were identified, viz., one in the vicinity of the percolation threshold, and two for high densities.

The percolation of rods and the effects of rod alignment on the electrical anisotropy of systems have been examined in several experimental and simulation works~\cite{Whitehouse1991,Lee1995,Dani1996,Du2005,Berhan2007,De2010,White2009,Ackermann2016}. For systems with aligned conductive fibers, different behaviors of electrical conductivity and different percolation thresholds were observed in the longitudinal and transverse directions~\cite{Whitehouse1991,Lee1995}. Injection-molded polymer composites filled with carbon-fibers demonstrated electrical anisotropy along different principal directions~\cite{Dani1996}. The observed behavior reflected a preferred axial orientation of fibers within the samples. The effect of carbon nanotube alignment on the percolation conductivity of polymer composites has been studied both experimentally and by means of MC simulations~\cite{Du2005}. It has been demonstrated that percolation conductivity depends on both alignment and concentration. Moreover, the highest conductivity was observed for slightly aligned, rather than isotropic systems. When particles are strongly aligned, the film resistivity becomes highly dependent on the measurement direction~\cite{Behnam2007}. MC simulation of partially oriented carbon nanotubes also revealed the effects of alignment on electrical conductivity~\cite{Berahman2013}. The electrical conductivity of percolated networks of rods with various aspect ratios and degrees of axial alignment has been simulated~\cite{White2009}. For anisotropic systems, a particular type of angular distribution produced by interaction with a flow has been considered. At a fixed volume fraction and aspect ratio, the simulated electrical conductivity decreased dramatically once a critical degree of orientation had been reached. Electrical percolation in quasi-two-dimensional metal nanowire networks has been simulated~\cite{Mutiso2013}. The electrical conductivity was evaluated by assuming that all the electrical resistance resulted from contact resistance at the rod-rod junctions. The preconditioned conjugate gradient iterative method was used to solve a system of linear Kirchhoff's  equations. The data were presented as a calculated plot of the optical transmittance versus the sheet resistance. The conditions for both optimal sheet resistance and optical transmittance were formulated. For ultra-transparent electrodes based on aligned silver rods significant electrical anisotropy was observed for films with high transmission ($>99$ \%)~\cite{Ackermann2016}. For films with lower transmission, the electrical anisotropy becomes negligible. This phenomenon was attributed to the effects of the junctions between the rods.

However, no detailed study on the relationship between anisotropy in electrical conductivity and the optical transparency of a two-dimensional film of partially oriented conducting rods has previously been performed. In this work, we consider continuous models of 2D systems where the rods have zero-width (i.e. an infinite aspect ratio) and they may intersect each other. For calculation of the filling fraction and electrical conductivity a discretization algorithm based on the use of a supporting mesh, viz., a square lattice of size $m\times m$ ($m = 64-2048$) was applied. The discretization of the structure of the infinitely thin rods is equivalent to the substitution of rods with polyominoes (e.g., see~\cite{Golomb1994}). The cells of supporting lattice covered by rods were assumed to be nontransparent and electrically conducting. The Frank--Lobb algorithm was applied to evaluate the electrical conductivity.

The rest of the paper is constructed as follows. In Section~\ref{sec:methods}, the technical details of the simulations are described, all necessary quantities are defined, and some test results are shown. Section~\ref{sec:results} presents our principal findings. Section~\ref{sec:conclusion} summarizes the main results.

\section{Methods}\label{sec:methods}
\subsection{Generation of 2D system of rodlike particles onto a plane}
Rods of length $l$ and zero thickness, $d=0$, (i.e., with an infinite aspect ratio, $a=l/d=\infty$) were deposited onto a plane substrate with a desired number density, $n$.
Newly-deposited particles were allowed to overlap previously deposited ones. Periodic boundary conditions were assumed, i.e., the rods were deposited onto a torus.

The rods were aligned with respect to a selected direction, $x$. To characterize the anisotropy, we used the mean order parameter defined as (see, e.g,~\cite{Frenkel1985})
\begin{equation}\label{eq:S}
s  = 2\langle \cos^2\theta \rangle - 1.
\end{equation}

Two distinct possibilities were taken into account for the anisotropic deposition of particles.
Model I was equivalent to that considered in~\cite{Balberg1983,Balberg1983a,Balberg1984};
the rods being randomly distributed within the interval $-\theta_m < \theta < \theta_m$, i.e., only some angles were allowed with equal probabilities~\cite{Balberg1983a}. In this case,
\begin{equation}\label{eq:stheta1}
   s  = \frac{2}{\theta_m}\int_{0}^{\theta_m} \cos^2 \theta \, d\theta -1 = \frac{\sin 2\theta_m} {2\theta_m}.
\end{equation}

The order parameter, $s$, defined as~\eqref{eq:S}, and the degree of anisotropy,
$A$, defined as~\eqref{eq:Anisotropy}, are related to each other as
\begin{equation}\label{eq:svsPP}
  s = \frac{2 A\left(A^2-1\right)}{\left(A^2+1\right)^2\arctan\left(2 A/\left(A^2-1\right)\right)}.
\end{equation}

In Model II, the orientations of the rods were distributed according to a normal distribution, i.e., all angles, $\theta$, were allowed, but with different probabilities. In this latter case,
\begin{equation}\label{eq:sII}
  s = \frac{2}{\sqrt{2\pi\sigma^2} } \int_{-\infty}^{\infty}  \exp\left( -\frac{\theta^2}{2\sigma^2} \right) \cos^2 \theta \,d \theta -1.
\end{equation}
The variance of the normal distribution ($\sigma^2 = \Var(\theta)$) is related to the desired mean order parameter as
\begin{equation}\label{eq:variance}
  \Var(\theta) = -0.5 \ln  s.
\end{equation}
%The order parameter and the degree of anisotropy are related to each other as
%\begin{equation}\label{eq:svsPPII}
%  A = \left( \erfi \left(  \sqrt{ - \ln ( s )}/2 \right)  \right) ^{-1}, s \to 1,
%\end{equation}
%where $\erfi$ is the imaginary error function (see, e.g.,~\cite{Abramowitz1974}).

\begin{figure}[htbp]
  \centering
\includegraphics[width=0.9\linewidth]{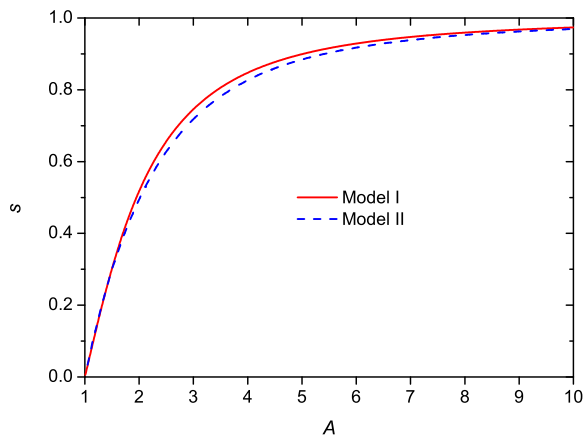}
  \caption{Order parameter, $s$, versus the degree of anisotropy, $A$. Model I calculated using~\eqref{eq:svsPP}, Model II calculated numerically.\label{fig:C0O_OrderParameter}}
\end{figure}

Figure~\ref{fig:C0O_OrderParameter} shows the order parameter, $s$, versus the degree of anisotropy, $A$, for both models.

Obviously, both the models are completely equivalent for the two limiting cases $s=0$ and $s=1$. Nevertheless, they demonstrate slightly different behaviors for intermediate values of $s$.

\subsection{Computation of the electrical conductivity}
\begin{figure}[htbp]
  \centering
  \includegraphics[width=0.9\linewidth]{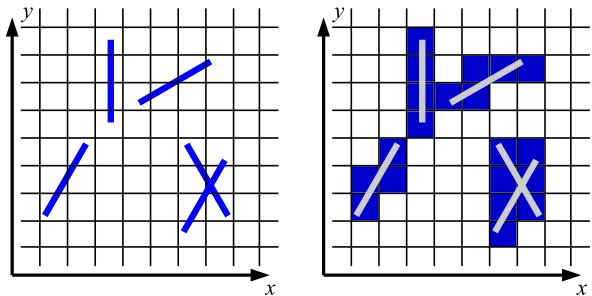}
  \caption{Transformation of the continuous problem of rods into a ``zoo of lattice animals'' (linear polyominoes).\label{fig:animals}}
\end{figure}

To take account of the electrical properties of both the low conductive substrate and highly conductive fillers, a discretization approach, involving a supporting mesh, viz., a square lattice of size $m\times m$ ($m=64$--2048) was used. When any part of a rod was situated inside a face (cell) of the supporting mesh, this cell was treated as nontransparent and conducting, otherwise the cell was assumed to be transparent and insulating. Such a discretization generated a ``zoo of lattice animals'', i.e., a set of polyominoes of different shapes and sizes, especially, for smaller values of $m$ (see Fig.~\ref{fig:animals}). This set is not complete because discretization of a rod cannot produce polyominoes of all possible shapes but only the so-called linear polyominoes, i.e., those polyominoes possessing the property that a line can be drawn that intersects the interior of every square in the polyomino~\cite{Friedman2001}.

\begin{figure}[htbp]
  \centering
  \includegraphics[width=0.9\linewidth]{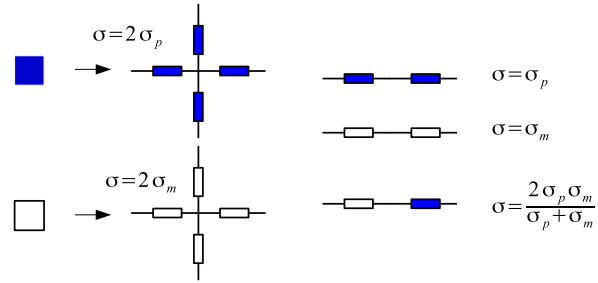}
  \caption{Transformation of the supporting mesh into a resistor network. All possible combinations of the conductivities are indicated.\label{fig:transformation}}
\end{figure}

Each occupied cell (a face of the lattice) was treated as conducting. To transform the mesh into a random resistor network (RRN), each cell was associated with a set of 4 conductors (Fig.~\ref{fig:transformation}).
The electrical conductivity of the particles $\sigma_p$  was supposed to be much larger than the electrical conductivity of the substrate $\sigma_m$, i.e., the electrical conductivity contrast $\Delta=\sigma_p/\sigma_m \gg 1$. We put $\sigma_m =1$, and $\sigma_p = 10^6$ in arbitrary units. Different electrical conductivities corresponding to the empty cells, $\sigma_m$, occupied cells, $\sigma_p$,  and between empty and occupied cells, $\sigma_{pm}=2\sigma_p \sigma_m / (\sigma_p+\sigma_m)$ were assumed (Fig.~\ref{fig:transformation}). In our calculations, the torus was unrolled, the two conducting buses were applied to the opposite borders of the lattice, and the electrical conductivity was calculated between these buses in the horizontal, $\sigma_x$ and vertical $\sigma_y$ directions (see~\cite{Lebovka2016,Lebovka2017} for the detail).

The Frank--Lobb algorithm was applied to calculate the electrical conductivity~\cite{Frank1988}.
For a quantitative description of the anisotropy of the electrical conductivity in the longitudinal and transversal directions, the electrical conductivity ratio, defined from the electrical contrast~$\Delta$,
\begin{equation}\label{eq:deltaxy}
\sigma_y /\sigma_x =\Delta^\delta,
\end{equation}
was used~\cite{Tarasevich2016}. $\delta=0$ for isotropic systems and $\delta\approx 1$ for highly anisotropic systems with
$\sigma_y /\sigma_x \approx \Delta$.

To characterize the insulator--conductor phase transition, we used the mean geometric value
\begin{equation}\label{eq:sigmag}
  \sigma_g =\sqrt{\sigma_m \sigma_p},
\end{equation}
and treated a system with conductivity $\sigma > \sigma_g$ as conducting, whilst a system with a conductivity $\sigma < \sigma_g$ was considered to be insulating. Note that the mean geometric conductivity  corresponds exactly to the prediction for the percolation threshold in the case of 2D systems with equal concentrations of the phases $p_c=1/2$ when both phases are in geometrically equivalent conditions (on average)~\cite{Dykhne1971}.

The use of the supporting mesh for the calculation of electrical conductivity is equivalent to the discretization of the structure of  the infinitely thin rods and the substitution of  them by anisotropic particles with a finite aspect ratio of the order of $k^*= m/L$.

\begin{figure}[htbp]
  \centering
    \includegraphics[width=0.9\linewidth]{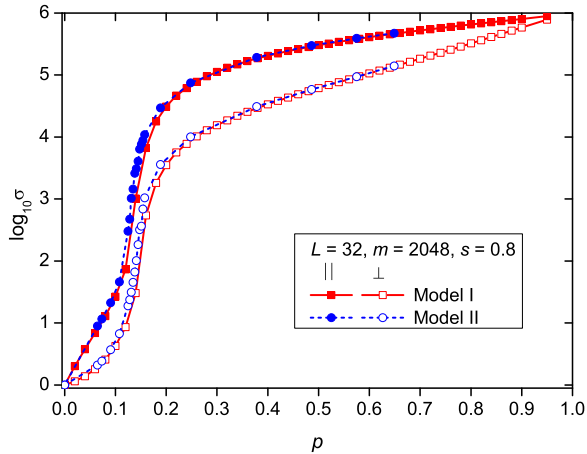}\\
  \caption{Comparison of the results obtained within the frameworks of Models I and II. Electrical conductivity, $\sigma$,  versus the filling fraction, $p$, for $m=2048$ and $s=0.8$. Here, the filled and open symbols correspond to the longitudinal and transversal directions, respectively. The lines are provided simply as visual guides. Solid lines correspond to Model I, dashed lines correspond to Model II.\label{fig:compare}}
\end{figure}

Although Model II looks a little bit more realistic, Model I has previously been widely used.​   This is the reason why all our computations were performed within the frameworks of both models.
Figure~\ref{fig:compare} presents examples of comparison of the electrical conductivity, $\sigma$, versus the filling fraction, $p$, for both models. We have found that differences in the results are of the same order as the statistical errors. Therefore, in this article we have presented all the results only for Model~II. Throughout the text, the error bars in the figures correspond to the standard deviations of the means. When not shown explicitly, they are of the order of the marker size.

The filling fraction (optical absorbance of the 2D film), $p$, defined as the number of occupied cells divided by the total number of cells (i.e., $m^2$) was used to characterize the system after discretization. There is no one-to-one correspondence between the number concentration of rods $n$ and the filling fraction $p$, i.e., one cannot predict the exact value of the packing density even when the number of deposited rods is known.

In all calculations, the quantities under consideration were typically averaged over 100 independent statistical runs for $m\leq256$, over 25 for $m=512$, over 10 for $m=1024$, and over 5 for $m=2048$, unless otherwise explicitly specified in the text.

\subsection{Scaling analysis}\label{app:scaling}
\begin{figure}[hbtp]
  \centering
  \includegraphics[width=0.9\linewidth]{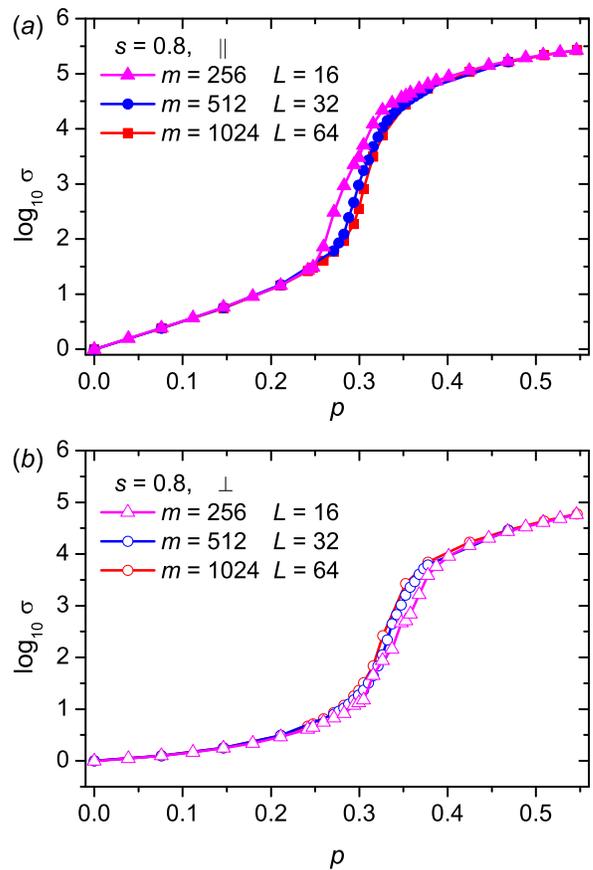}
  \caption{Example of scaling analysis for electrical conductivity, $\sigma$ vs filling fraction, $p$, for the different values of $L= 16, 32, 64$ but with a fixed value $m/L=16$. (a) longitudinal conductivity, (b)~transversal conductivity. The lines are provided simply as visual guides.\label{fig:scalingp}}
\end{figure}

The size of the system was $L\times L$  and, in the present work, all calculations were performed using $L = 32l$. Our choice is based on a scaling analysis. We performed calculation of the electrical conductivities for regions of different sizes $L=16,32,64$. Figure~\ref{fig:scalingp} compares the results for a fixed value of $k^*$. The differences in the electrical conductivity in the vicinity of zones of the insulator-to-conductor transition were insignificant, and were almost completely absent outside of these regions.

\section{Results}\label{sec:results}
\subsection{Polyominoes formed by  discretization}

In practice, the discretization transformed the continuous problem of rods into a lattice problem of linear polyominoes, i.e particles having different shapes and length. The mean number of cells, $k$, in such a polyomino depends on the angle $\theta$ between the axis of the rod and the selected direction, $x$
\begin{equation}\label{eq:angledistrib}
\frac{k}{k^*} = \frac{1}{k^*} +  \sqrt{2} \sin\left( \theta + \frac{\pi}{4}\right), \quad 0 \leqslant \theta \leqslant \frac{\pi}{4},
\end{equation}
where $k^* = m/L$ (see Appendix~\ref{app:distribution} for details).

Figure~\ref{fig:angledistrib} demonstrates how the size of a polyomino depends on the angle $\theta$. In computer simulation, a rod was randomly placed onto the plane with the angle $\theta = 0$ to the axis $x$ and the size of the generated polyomino was calculated. Then, the rod was turned to an angle equal to $1^\circ$ and the size of the generated polyomino calculated once again. These turns continued until the angle reached $\theta=90^\circ$. After that, the results for each angle were averaged over $10^5$ independent placements of the rods.  The dependence obtained by means of this simulation is indistinguishable from~\eqref{eq:angledistrib}.
\begin{figure}[htbp]
  \centering
  \includegraphics[width=0.9\linewidth]{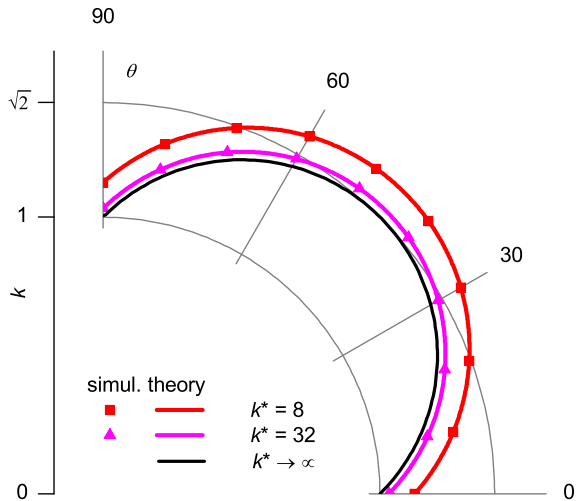}
  \caption{Examples of the size distribution of ``lattice animals'' (polyominoes) vs angle for different values of $m$ in the polar diagram. The results were averaged over $10^5$ independently located rods. $k^* = m/L$. Only each tenth point is shown to allow a clearer view. The curves correspond to Eq.~\eqref{eq:angledistrib}.
  \label{fig:angledistrib}}
\end{figure}

\begin{figure}[htbp]
  \centering
  \includegraphics[width=0.9\linewidth]{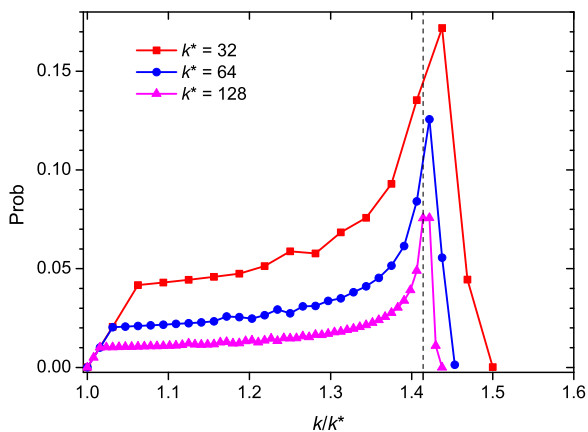}
  \caption{Examples of the distribution of ``lattice animals'' (polyominoes), by size, in an  isotropic system ($s=0$) for three different values of $m$ according to Eq.~\eqref{eq:Psize}. The solid lines are provided simply as visual guides. The vertical dashed line corresponds to $\sqrt2$.\label{fig:sizedistribution}}
\end{figure}

Figure~\ref{fig:sizedistribution} shows  the  distribution of polyominoes, by size, for different values of $m$ calculated according to the analytical expression~\eqref{eq:Psize}. For instance, when $k^*=128$, the possible sizes of polyominoes vary from 128 to 309.  The probabilities are exact, and such non-monotonic behavior is an intrinsic property, not a computational error.

It is noticeable, that dispersity and polymorphism are inherent properties of a discrete system produced by means of discretization from the continuous system. They never vanish even when $m \to \infty$. This evidences that the properties of such a discrete system are related, but not identical, to the properties of the original continuous system.

\subsection{Electrical conductivity behavior}

\begin{figure}[htbp]
  \centering
  \includegraphics[width=0.9\linewidth]{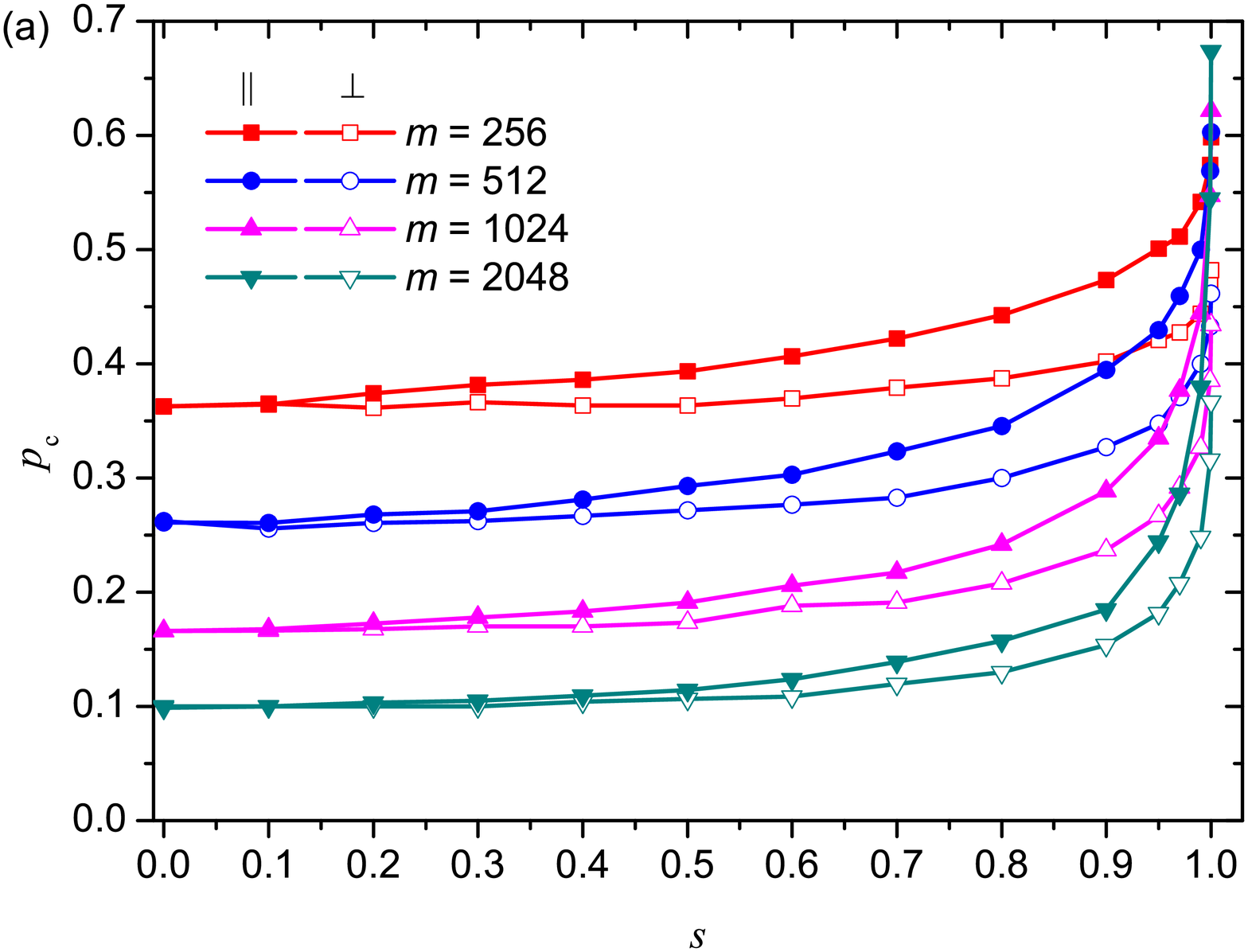}\\
  \includegraphics[width=0.9\linewidth]{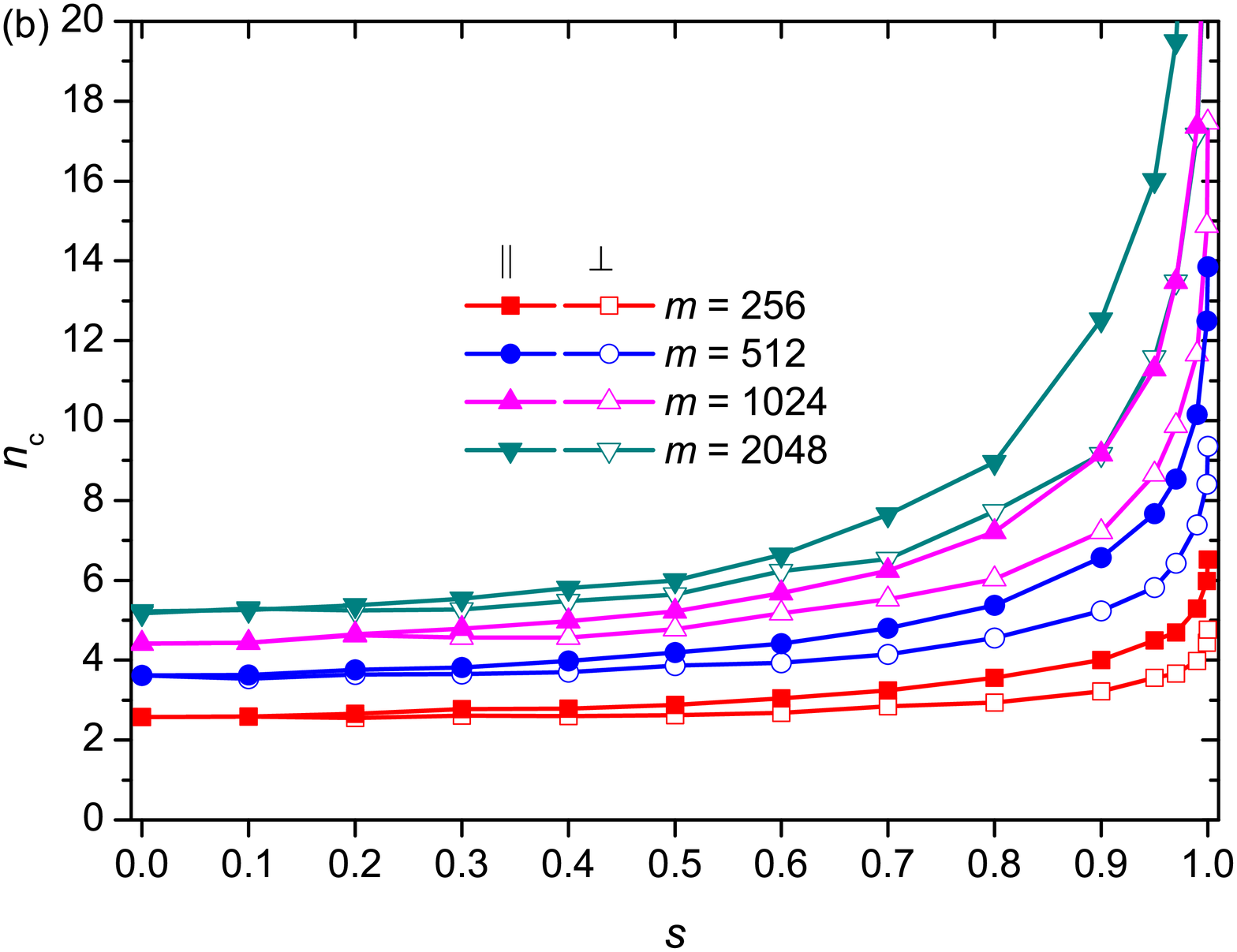}\\
\caption{Critical values of the filling ratio, $p_c$(a) and the number density, $n_c$ (b), versus the order parameter, $s$, at different values of $m$. Here, the filled and open symbols correspond to the longitudinal and transversal directions, respectively. The lines are provided simply as visual guides.\label{fig:II_vs_s}}
\end{figure}

Figure~\ref{fig:II_vs_s} presents the dependencies of the critical values of the filling ratio, $p_c$(a) and number density, $n_c$ (b), versus the order parameter, $s$, at different values of $m$. These values of $p_c$ and $n_c$ in the longitudinal direction (filled symbols) always exceeded those in the transversal direction (open symbols) and the rate of growth of these with increase of order parameter $s$. Moreover, an increase of $m$ resulted in a decrease of $p_c$ and an increase in $n_c$.

\begin{figure}[htbp]
  \centering
\includegraphics[width=0.9\linewidth]{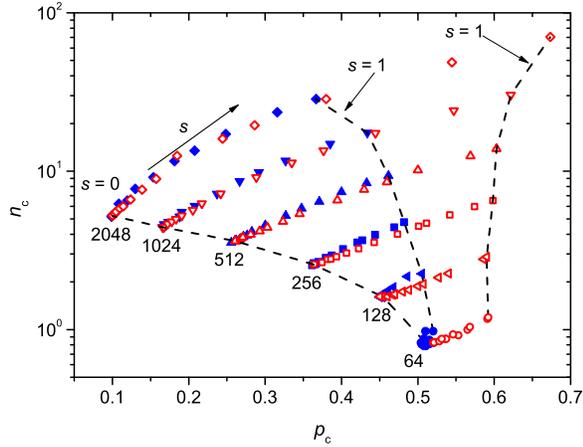}\\
\caption{Critical number density, $n_c$, versus the critical filling ratio, $p_c$ at different values of $m$ and the order parameter, $s$. Here, the filled and open symbols correspond to the longitudinal and transversal directions, respectively. The lines are provided simply as visual guides.
\label{fig:II_pc_vs_nc}}
\end{figure}

The correlations between critical values of the filling ratio, $p_c$, and number density, $n_c$, presented in Fig.~\ref{fig:II_pc_vs_nc} evidence the quite different behavior of completely ordered ($s=1$) and partially disordered ($s\neq 1$) systems.
When $m \to \infty$ (infinite discretization), for completely ordered systems ($s=1$), we observed an infinite increase in value of $n_c$ in the longitudinal direction, whereas such increase was finite  in the transversal direction (see dashed lines in Fig.~\ref{fig:II_pc_vs_nc}). For partially disordered ($s\neq 1$) systems, such increase was always finite in both directions.

\begin{figure}[htbp]
  \centering
  \includegraphics[width=0.9\linewidth]{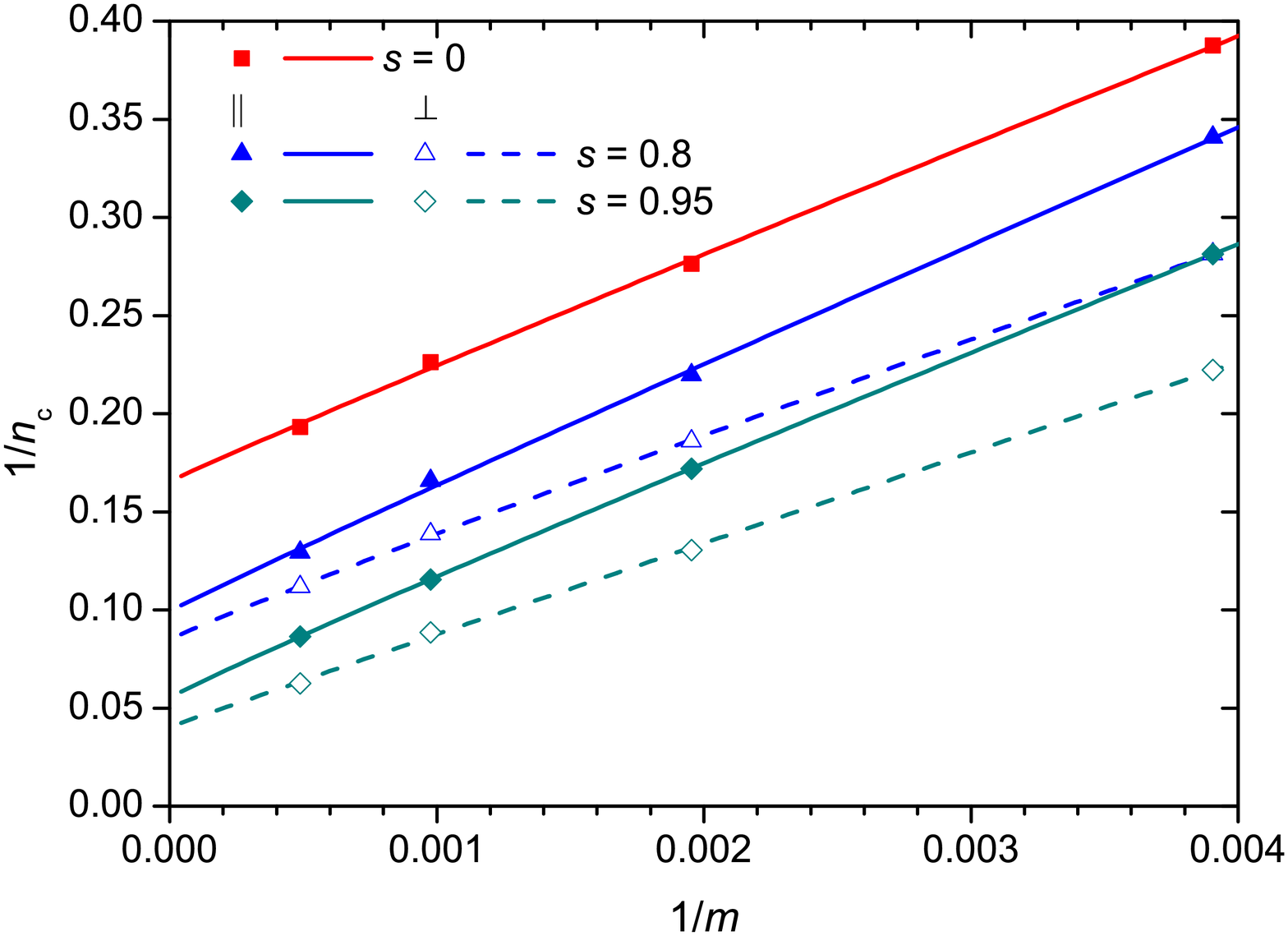}\\
\caption{Examples of fittings of $1/n_c$ versus $1/m$ with the power equation $1/n_c(m)=1/n_c(\infty)+\alpha/m^\beta$ used for evaluation of the values of $n_c(\infty)$. Here, $\alpha$ and $\beta$ are the fitting parameters.
\label{fig:C04_fitting}}
\end{figure}

The data obtained from the $p_c(m)$ dependencies were used for fitting with the power equation $1/n_c(m)=1/n_c(\infty)+\alpha/m^\beta$ and for evaluation of the values of $n_c(\infty)$. The examples of such fittings are presented in Fig.~\ref{fig:C04_fitting}.

\begin{figure}[htbp]
  \centering
\includegraphics[width=0.9\linewidth]{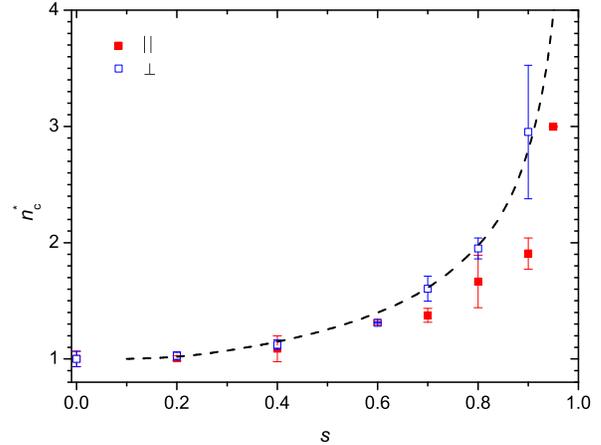}\\
\caption{Relative critical number density, $n^*=n_c/n_c^i$, ($n_c^i \approx 6.044$) versus order parameter, $s$, for the values obtained in the limit of $m\rightarrow\infty$. Here, the filled and open symbols correspond to the longitudinal and transversal directions, respectively. The dashed line was obtained using Eq.~\eqref{eq:NumberDensityAnisotropy}.
\label{C05_nr_vs_s}}
\end{figure}

Figure~\ref{C05_nr_vs_s} presents the relative critical number density, $n^*=n_c/n_c^i$, versus the order parameter, $s$. Here, the values were obtained in the limit of $m\rightarrow\infty$ and where $n_c^i$ corresponds to the isotropic case, $s=0$. The dashed line was obtained using the approximation derived in~\cite{Balberg1983a,Balberg1984a}. Note that the estimated value of $n^c_i\approx 6.044$ noticeably exceeds that obtained for the zero-width rods, e.g., $n^c_i\approx 5.64$~\cite{Li2009,Mertens2012}. This evidently reflects the transformation of  the zero-width rods to polyominoes of different shapes and sizes. Significant anisotropy in the $n^*$ values was observed and the calculated values in the longitudinal direction were very close to the approximation of Eq.~\eqref{eq:NumberDensityAnisotropy}.

\begin{figure}[htbp]
  \centering
\includegraphics[width=0.9\linewidth]{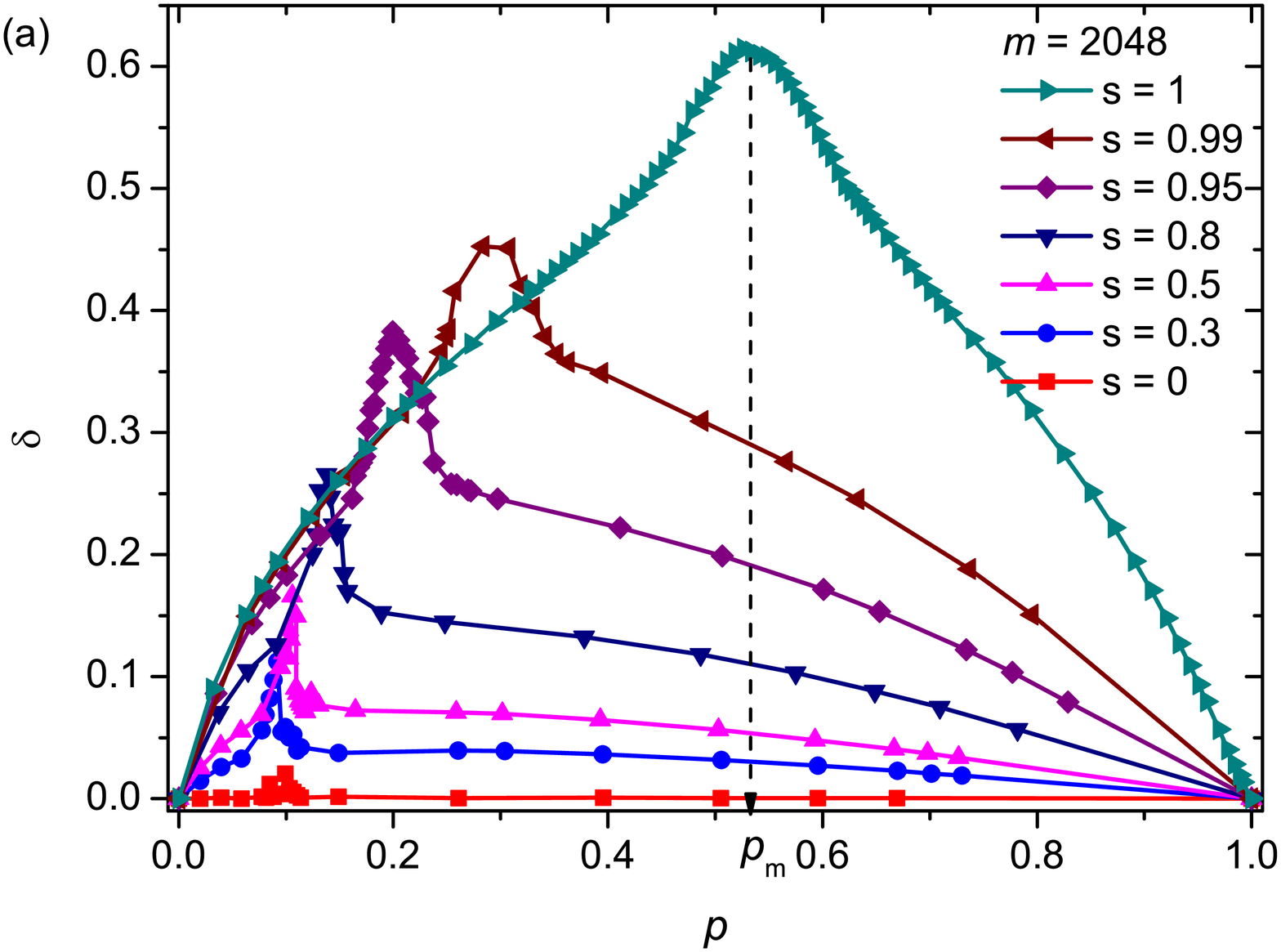}\\
\includegraphics[width=0.9\linewidth]{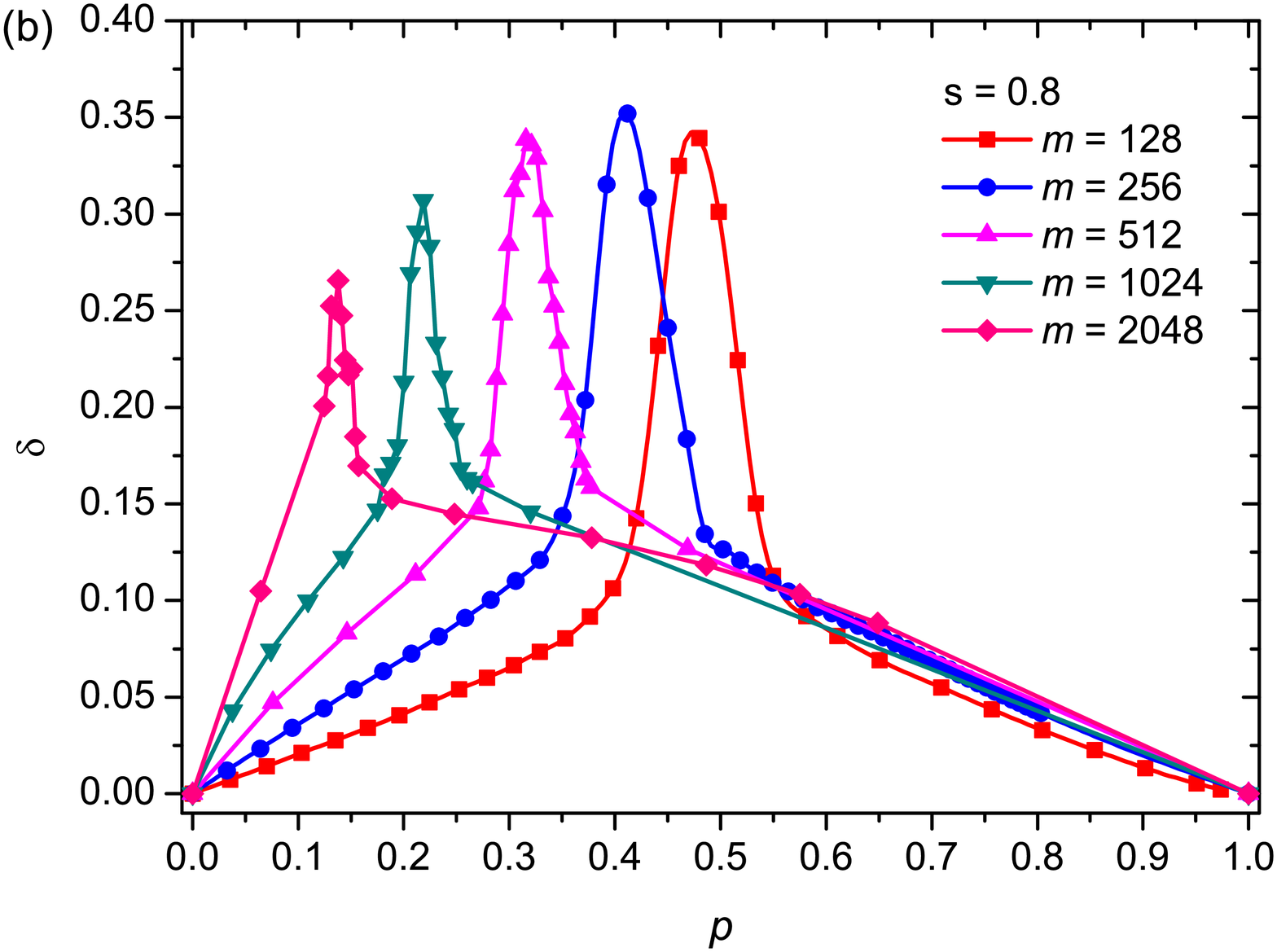}\\
\includegraphics[width=0.9\linewidth]{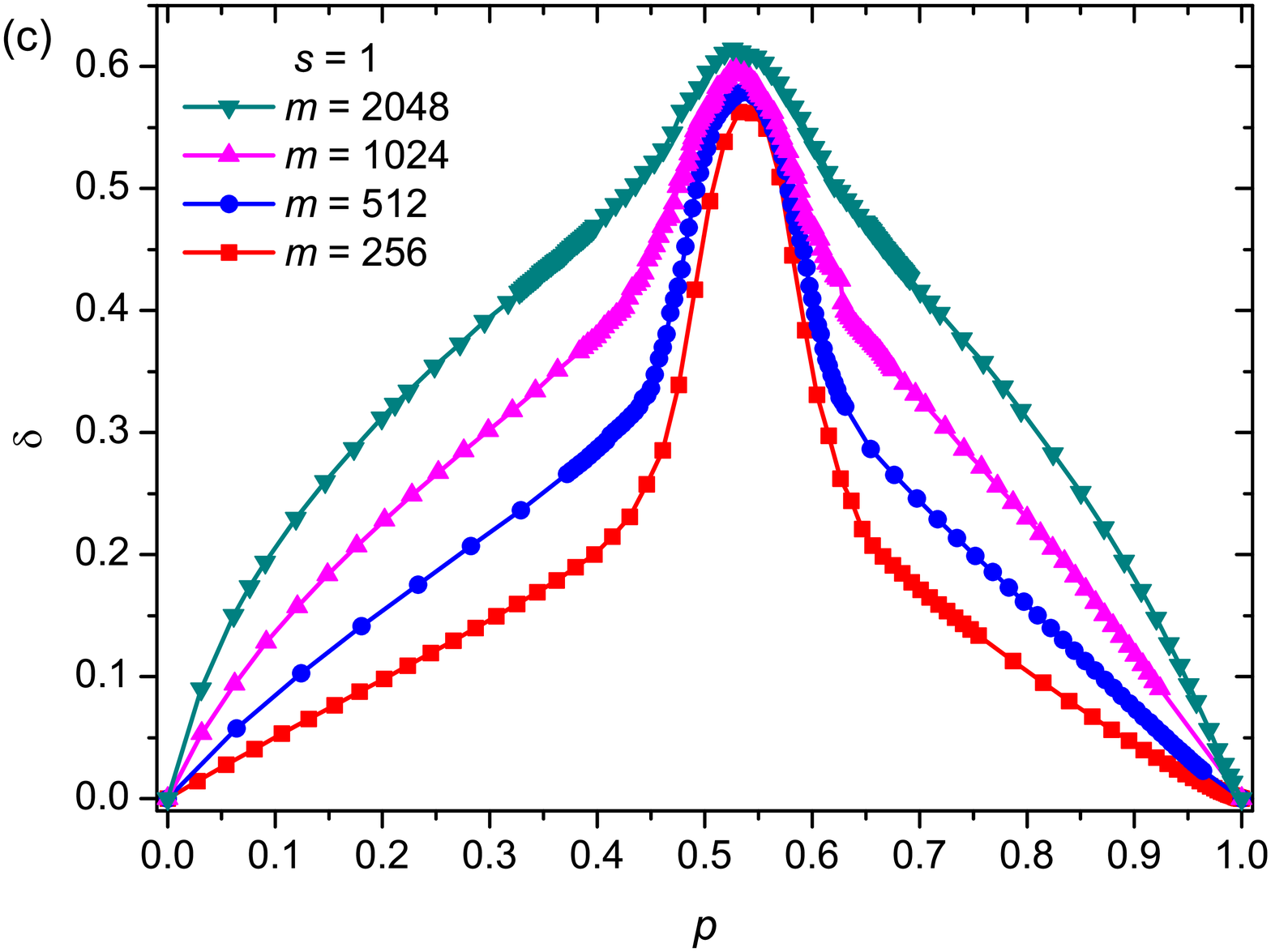}\\
\caption{Electrical anisotropy ratio, $\delta$, versus the filling fraction, $p$, (a) for $m=2048$ and different values of order parameter $s$, (b) for fixed value of order parameter, $s=0.8$ and different values of $m$, (c)  for fixed value of order parameter,  $s=1$, and different values of $m$.\label{II_delta_vs_p}}
\end{figure}

Figure~\ref{II_delta_vs_p} presents examples of the electrical anisotropy ratio, $\delta$, versus the filling fraction, $p$, for a fixed value of $m=2048$ and for different values of the order parameter, $s$ (Fig.~\ref{II_delta_vs_p}(a)), as well as for two values of the order parameter, $s=0.8$ and $s=1$ and different values of $m$ (Fig.~\ref{II_delta_vs_p}(b),(c)). It is remarkable that the position of the $\delta(p)$ maximum at $p_m$ was almost independent of $m$ for an ideally oriented system (Fig.~\ref{II_delta_vs_p}(c)). However, the value of $p_m$ significantly decreased with increasing $m$ (increase in the effective aspect ratio of the rods) for $s<1$ (see, e.g., Fig.~\ref{II_delta_vs_p}(c) for $s=0.8$). For the given $m$ (see, e.g., $m=2048$ in Fig.~\ref{II_delta_vs_p}(a)) the value of $p_m$ decreased with decrease of the order parameter, while significant electrical anisotropy was observed  for $s<1$. For example, for $s=0.5$  the anisotropy value was $\delta \approx 0.18$ at $p_m \approx 0.1$. Note that we can expect even more complexity in the presence of rod-rod junction resistances~\cite{Ackermann2016}. So, the alignment of the rods can significantly affect the electrical anisotropy and optical transmission in their dependence on the values of $s$ and $m$. This corresponds to the possibility of preparing an of optically transparent system with small values of $p_m$ and a high electrical anisotropy using rods with a high aspect ratio.

%\clearpage
\section{Conclusion}\label{sec:conclusion}

We have simulated the electrical properties of 2D films composed of a low conductive substrate (host matrix) and highly conductive zero-width rods (fillers). The rods were randomly deposited onto a plane substrate with overlapping being allowed. Moreover, they were aligned with respect to a selected direction and to a desired order parameter, $s$.  To evaluate the filling fraction (or non-transparency), $p$, and the electrical conductivity, $\sigma$, a discretization approach  involving a supporting square mesh was used. This approach allowed assessment of the electrical conductivities both of the insulating (low conductive) host matrix and high conducting filler particles. As a result of discretization, the zero-width conducting rods are transformed into conducting polyominoes of different shapes and sizes. It was demonstrated that the polymorphism and dispersity of the polyominoes could not be eliminated even at infinitely fine discretization. For example, for a disordered system ($s=0$) the estimated critical number density for the polyominoes $n_c^i \approx 6$ was greater than that for zero-width rods, $n_c^i =5.6372858(6)$~\cite{Mertens2012}. Alignment of the fillers could result in noticeable electrical anisotropy. The electrical properties of the films with perfectly aligned rods ($s = 1$) were different from that for slightly disordered systems ($0.6< s < 1$). For a perfectly aligned system, the highest film anisotropy was observed at high filler content ($p \approx 0.55$) for essentially non-transparent films. For non-perfectly aligned systems, high electrical anisotropy was observed at smaller filler content and this suggests a potential method for the production of optically transparent and highly anisotropic conducting films.

%\clearpage
\begin{acknowledgments}
We acknowledge funding provided by the National Academy of Sciences of Ukraine, Project No.~43/18-H and 15F (0117U006352) (N.I.L. and N.V.V.) and the Ministry of Education and Science of the Russian Federation, Project No.~3.959.2017/4.6 (Yu.Yu.T., A.V.E., and R.K.A.). The authors would like to thank V.V.~Chirkova for her technical assistance.
\end{acknowledgments}

\appendix
\section{Size distribution of polyominoes}\label{app:distribution}
\begin{figure}[htb]
  \centering
  \includegraphics[width=\linewidth]{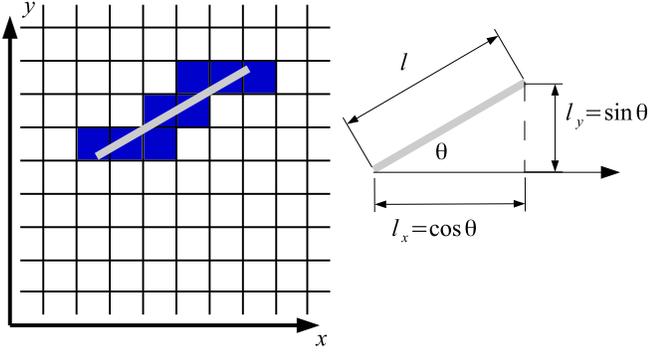}
  \caption{Example of a polygon formed by a line segment on a checkered paper. The size of the polyomino is 8.\label{fig:needle}}
\end{figure}

Let there be a line segment (a rod) of length $l$, which is randomly placed on a sheet of checkered paper with cells of unit size $1 \times 1$.  All cells in which there is at least some part of the linear segment are considered as belonging to the polyomino (Fig.~\ref{fig:needle}).

\begin{figure}[htb]
  \centering
  \includegraphics[width=0.7\linewidth]{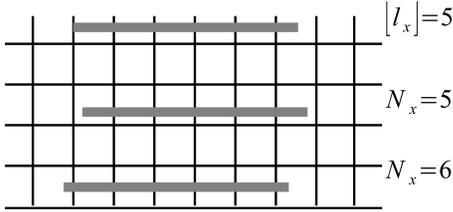}
  \caption{Example of a horizontal line segment on checkered paper, $5<l<6$, i.e., $\lfloor l_x \rfloor =5$. The line segment may intersect only 5 or 6 vertical lines.\label{fig:intersections}}
\end{figure}

Where the numbers of intersections with horizontal and vertical lines by a line segment are denoted as $N_y, N_x$, respectively, only two possibilities can be realized, viz, $N_i = \lfloor l_i\rfloor$ or $N_i = \lfloor l_i\rfloor + 1$, where $i=x,y$  and $\lfloor \cdot \rfloor$ means the floor function, i.e. the function that takes as its input a real number and gives as an output the greatest integer less than or equal to the input number (Fig.~\ref{fig:intersections}).

Thus, the probabilities, $P$, that a line segment intersects the given number of vertical (horizontal) lines are
\begin{align}\label{eq:prob}
P\left( N_x = \lfloor l_x\rfloor + 1 \right) &= \left\{ l_x \right\} = \left\{ l \cos \theta \right\} = p_x,\\
P\left( N_x = \lfloor l_x\rfloor \right) &= 1 - \left\{ l_x \right\} = 1 - p_x = q_x,\\
P\left( N_y = \lfloor l_y\rfloor + 1 \right) &= \left\{ l_y \right\} = \left\{ l \sin \theta \right\} = p_y,\\
P\left( N_y = \lfloor l_y\rfloor \right) &= 1 - \left\{ l_y \right\} = 1 - p_y = q_y.
\end{align}
Here $l_x = l \cos \theta$, $l_y = l \sin \theta$, $\{\cdot\}$ means the fractional part of the number.

The number of cells in the polyomino is equal to
$$
N = N_x + N_y + 1.
$$
Since
$$
\min(N_x) + \min(N_y) + 1 \leq N \leq \max(N_x) + \max(N_y) + 1,
$$
only a few possibilities can be taken into account.
Hence, the probabilities of producing polyominoes of given sizes are
\begin{align}\label{eq:prob1}
P_1 &= P\left( N = \lfloor l_x\rfloor + \lfloor l_y\rfloor + 1 \right)= q_x q_y,
\\
P_2 &= P\left( N = \lfloor l_x\rfloor + \lfloor l_y\rfloor + 2 \right)= p_x q_y + q_x p_y,
\\
P_3 &= P\left( N = \lfloor l_x\rfloor + \lfloor l_y\rfloor + 3 \right)=  p_x p_y.
\end{align}

Thus, the probability that a rod with angle $\theta$ relating to the axis $x$ produces a $k$-omino equals
\begin{equation}\label{eq:probabsize}
  \Prob(k|\theta) =
  \begin{cases}
    p_xp_y, & \mbox{if } k = \lfloor l_x\rfloor + \lfloor l_y\rfloor + 3,\\
    p_x q_y + q_x p_y, & \mbox{if } k = \lfloor l_x\rfloor + \lfloor l_y\rfloor + 2, \\
    q_x q_y, & \mbox{if }  k = \lfloor l_x\rfloor + \lfloor l_y\rfloor + 1\\
    0, & \mbox{otherwise}.
  \end{cases}
\end{equation}

If all angles $\theta$ are equiprobable, then the probability of finding a $k$-omino can be obtained by the integration of $\Prob(k|\theta)$ over all possible angles
\begin{equation}\label{eq:Psize}
  \Prob(k) = \frac{4}{\pi} \int_0^{\pi/4} \Prob(k|\theta) \, d\theta.
\end{equation}

The mathematical expectation, i.e. mean size of the polyominoes, $k$, is
\begin{multline}\label{eq:MVdef}
k = \left( \lfloor l_x\rfloor + \lfloor l_y\rfloor + 3 \right) P_3 +\\
\left(  \lfloor l_x\rfloor + \lfloor l_y\rfloor + 2 \right) P_2 +
\left( \lfloor l_x\rfloor + \lfloor l_y\rfloor + 1 \right) P_1.
\end{multline}
Substitution of $P_1,P_2,P_3$ from~\eqref{eq:prob1} and $p_x,p_y,q_x,q_y$ from~\eqref{eq:prob} into~\eqref{eq:MVdef} after obvious transformations yields
\begin{equation}\label{eq:AD1}
k = 1 +  \sqrt{2} l \sin\left( \theta + \frac{\pi}{4}\right).
\end{equation}

In our work, the size of the line segment is an integer number $l = k^*$, where  $k^* = m/L$, hence,
\begin{equation}\label{eq:AD}
\frac{k}{k^*} %= \frac{1}{k^*} + \sin \theta + \cos \theta
= \frac{1}{k^*} +  \sqrt{2} \sin\left( \theta + \frac{\pi}{4}\right).
\end{equation}
The possible sizes of the polyominoes lie in the interval $ k \in \left[ k^*,( 1 +  \sqrt 2) k^*\right)$.

For the limiting case, when $k^* \gg 1$, formula~\eqref{eq:AD} simplifies to
\begin{equation}\label{eq:ADlim}
\frac{k}{k^*} %= \frac{1 + \tan \theta}{\sqrt{1 + \tan^2 \theta}}
= \sqrt{2} \sin\left( \theta + \frac{\pi}{4}\right).
\end{equation}

The probability density function of the mean polyomino size (PDF) is
\begin{equation}\label{eq:PDF}
  f(k) = \frac{4}{\pi \sqrt{ 2 - \left(k/k^* - {k^*}^{-1}\right)^2}}.
\end{equation}
For $k^* \gg 1$, the dispersion of possible sizes of the polyominoes generated by different rods with a fixed angle tends to zero, hence, \eqref{eq:PDF} may be treated as the size distribution of the polyominoes. The PDF has a singularity at $k = \sqrt2 k^*$. In this limiting case, the mean size of the polyominoes is $4 k^* / \pi$.

%\clearpage
\bibliography{Manuscript_Anisotropy}

\end{document}